\documentclass[12pt]{iopart}
\usepackage[dvips]{graphicx}
\begin{document}
%

\title{Flow in Au+Au Collisions at RHIC}

\author {Marguerite Belt Tonjes$^1$ for the PHOBOS Collaboration}
\noindent
B~B~Back$^2$,
M~D~Baker$^3$,
M~Ballintijn$^4$,
D~S~Barton$^3$,
R~R~Betts$^5$,
A~A~Bickley$^1$,
R~Bindel$^1$,
A~Budzanowski$^6$,
W~Busza$^4$,
A~Carroll$^3$,
M~P~Decowski$^4$,
E~Garc\'{\i}a$^5$,
N~George$^{2,3}$,
K~Gulbrandsen$^4$,
S~Gushue$^3$,
C~Halliwell$^5$,
J~Hamblen$^7$,
G~A~Heintzelman$^3$,
C~Henderson$^4$,
D~J~Hofman$^5$,
R~S~Hollis$^5$,
R~Ho\l y\'{n}ski$^6$,
B~Holzman$^3$,
A~Iordanova$^5$,
E~Johnson$^7$,
J~L~Kane$^4$,
J~Katzy$^{4,5}$,
N~Khan$^7$,
W~Kucewicz$^5$,
P~Kulinich$^4$,
C~M~Kuo$^8$,
W~T~Lin$^8$,
S~Manly$^7$,
D~McLeod$^5$,
A~C~Mignerey$^1$,
R~Nouicer$^5$,
A~Olszewski$^6$,
R~Pak$^3$,
I~C~Park$^7$,
H~Pernegger$^4$,
C~Reed$^4$,
L~P~Remsberg$^3$,
M~Reuter$^5$,
C~Roland$^4$,
G~Roland$^4$,
L~Rosenberg$^4$,
J~Sagerer$^5$,
P~Sarin$^4$,
P~Sawicki$^6$,
W~Skulski$^7$,
P~Steinberg$^3$,
G~S~F~Stephans$^4$,
A~Sukhanov$^3$,
J-L~Tang$^8$,
A~Trzupek$^6$,
C~Vale$^4$,
G~J~van~Nieuwenhuizen$^4$,
R~Verdier$^4$,
F~L~H~Wolfs$^7$,
B~Wosiek$^6$,
K~Wo\'{z}niak$^6$,
A~H~Wuosmaa$^2$ and
B~Wys\l ouch$^4$

\vspace{3mm}

\small
\noindent
$^1$~University of Maryland, College Park, MD 20742, USA\\
$^2$~Argonne National Laboratory, Argonne, IL 60439-4843, USA\\
$^3$~Brookhaven National Laboratory, Upton, NY 11973-5000, USA\\
$^4$~Massachusetts Institute of Technology, Cambridge, MA 02139-4307, USA\\
$^5$~University of Illinois at Chicago, Chicago, IL 60607-7059, USA\\
$^6$~Institute of Nuclear Physics, Krak\'{o}w, Poland\\
$^7$~University of Rochester, Rochester, NY 14627, USA \\
$^8$~National Central University, Chung-Li, Taiwan 

\ead{tonjes@bnl.gov}

\begin{abstract}
     The study of flow can provide information on the initial state dynamics and the degree of equilibration attained in heavy ion collisions. This contribution presents results for both elliptic and directed flow as determined from data recorded by the PHOBOS experiment in Au+Au runs at RHIC at $\sqrt{s_{_{NN}}} =$ 19.6, 130 and 200 GeV. The PHOBOS detector provides a unique coverage in pseudorapidity for measuring flow at RHIC. The systematic dependence of flow on pseudorapidity, transverse momentum, centrality and energy is
discussed.
\end{abstract}
\begin{figure}[ht]
\begin{center}
\includegraphics[height=6cm]{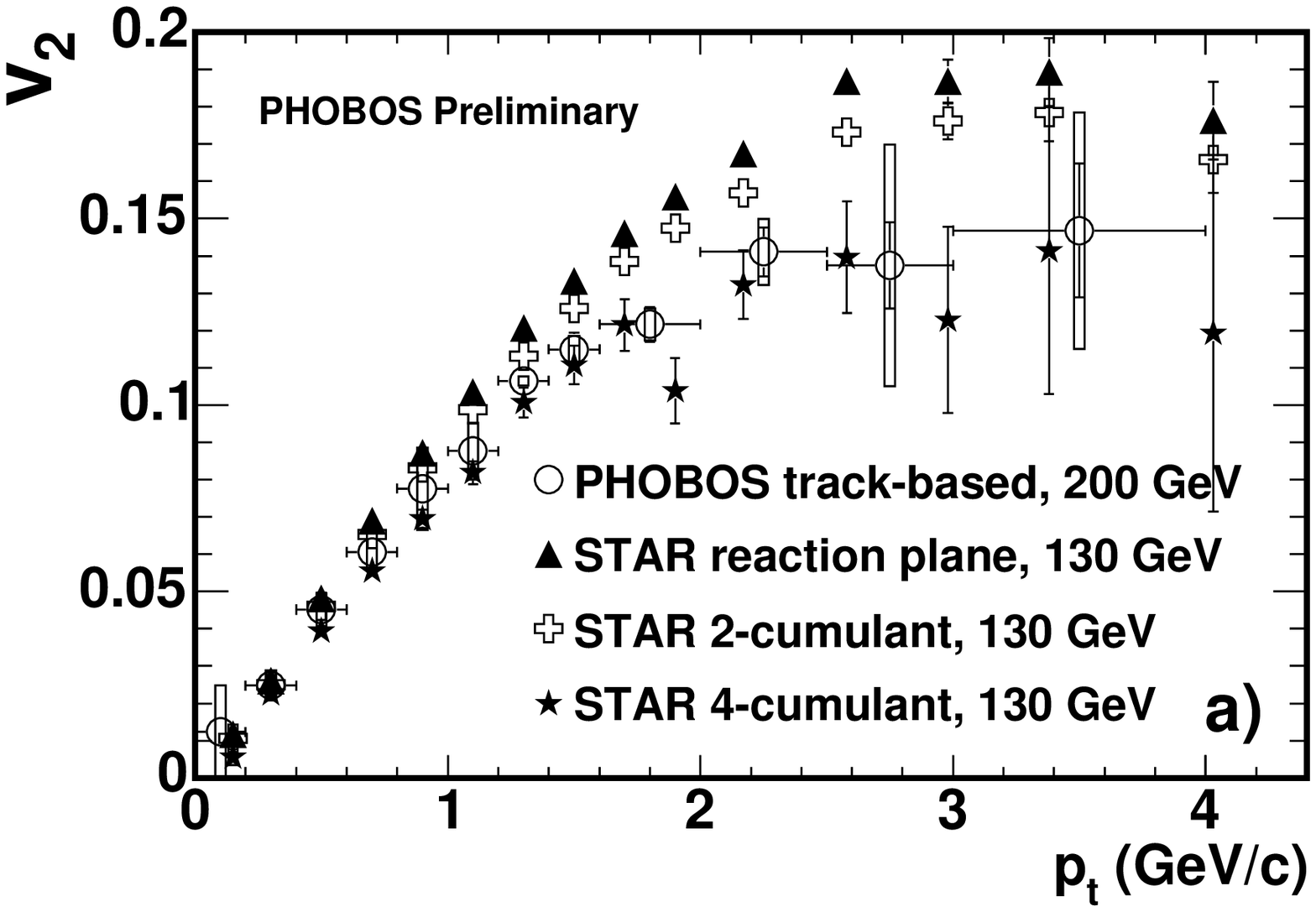}
\includegraphics[height=6cm]{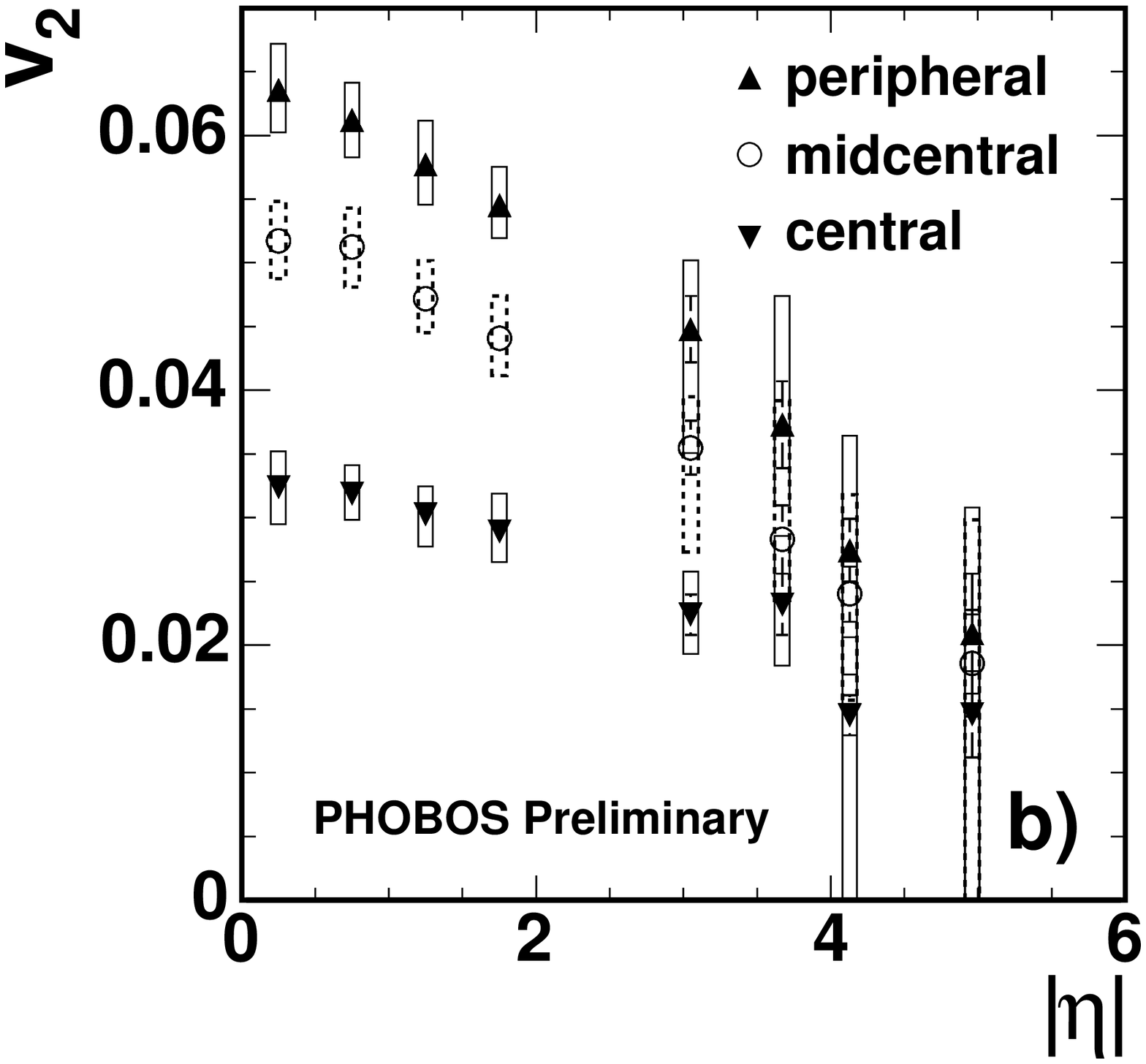}
\end{center}
\caption{\label{fig:v2} a) Comparison of PHOBOS (circles) $v_{2}$ for 0-55\% centrality Au+Au collisions to STAR Au+Au collisions at 5-53\% centrality. STAR results are shown for 3 methods: reaction plane (triangles), 2-particle cumulant (open crosses), and 4-particle cumulant (stars)\cite{starcumulant}. PHOBOS error bars are statistical, with error boxes representing systematic uncertainty at a 90\% confidence level(CL). b) $v_{2}$ vs. $|\eta|$ at different centralities: peripheral (25-50\%), midcentral(15-25\%), and central(3-15\%) for 200 GeV Au+Au collisions. Statistical error bars are smaller than the data points, and boxes show systematic errors at 90\% CL.}
\end{figure}

\vspace{-8mm}

\section{Introduction}
Compression occuring in the early stages of a relativistic heavy ion collision is believed to be responsible for anisotropic flow\cite{theory_ollitrault,theory_kolb}. The measurement of flow probes the thermalization of the system, and aids in understanding the evolution of the state of matter created. A spatial and momentum anisotropy is created that is most readily observed in the first ($v_{1}$ - directed flow) and second ($v_{2}$ - elliptic flow) Fourier coefficients of the azimuthal distributions measured with respect to the reaction plane of the collision. In this analysis, the reaction plane method and a subevent technique are used to calculate flow\cite{Poskanser}. 

\section{Detector}
The PHOBOS experiment measured collisions of Au nuclei at $\sqrt{s_{NN}}$ of 19.6, 130, and 200 GeV in 2000 and 2001\cite{phobos_det}. Silicon sub-detectors were used for flow analysis, including the octagon and ring detectors which cover $-5.4 < \eta < 5.4$ and all azimuthal angles. In addition, the spectrometer arms allow for charged particle momentum measurement in $0 < \eta < 1.5$. Two independent methods are employed in the flow studies: hit-based analysis using the octagon and rings\cite{phflow}, and track-based derivation using the spectrometer and octagon\cite{qm02}. Both methods allow for widely separated subevents which reduces non-flow effects. The susceptibility of the two methods to background effects is quite different due to the tracking, and good agreement has been shown between PHOBOS elliptic flow measurements with hit-based and track-based methods\cite{qm02}.

\section{Elliptic Flow}
The charged hadron $v_{2}$ as a function of $p_{t}$, measured by PHOBOS for 200 GeV Au+Au collisions is shown by the circles in figure \ref{fig:v2}a). Also shown are STAR measurements for 130 GeV Au+Au collisions analyzed with three different methods. Previous measurements demonstrate that elliptic flow changes little from 130 to 200 GeV at $p_{t}> 1$ GeV/c\cite{v2match}. The PHOBOS measurement matches STAR's 4-particle cumulant results, which has been shown to be insensitive to non-flow effects\cite{starcumulant}. This confirms our expectation that our flow results are not strongly biased by non-flow contributions.

For 200 GeV Au+Au collisions we study the pseudorapidity dependence of elliptic flow as a function of centrality. Results of the two methods agree and are combined and plotted as a function of $|\eta|$ in figure \ref{fig:v2}b) for three broad centrality ranges. The peripheral data do not appear to be flat, event at midrapidity. Within the uncertainties, the shape of $v_{2}(|\eta|)$ is not strongly centrality dependent, appearing to differ by only a scale factor.

\begin{figure}[ht]
\begin{center}
\includegraphics[height=8.1cm]{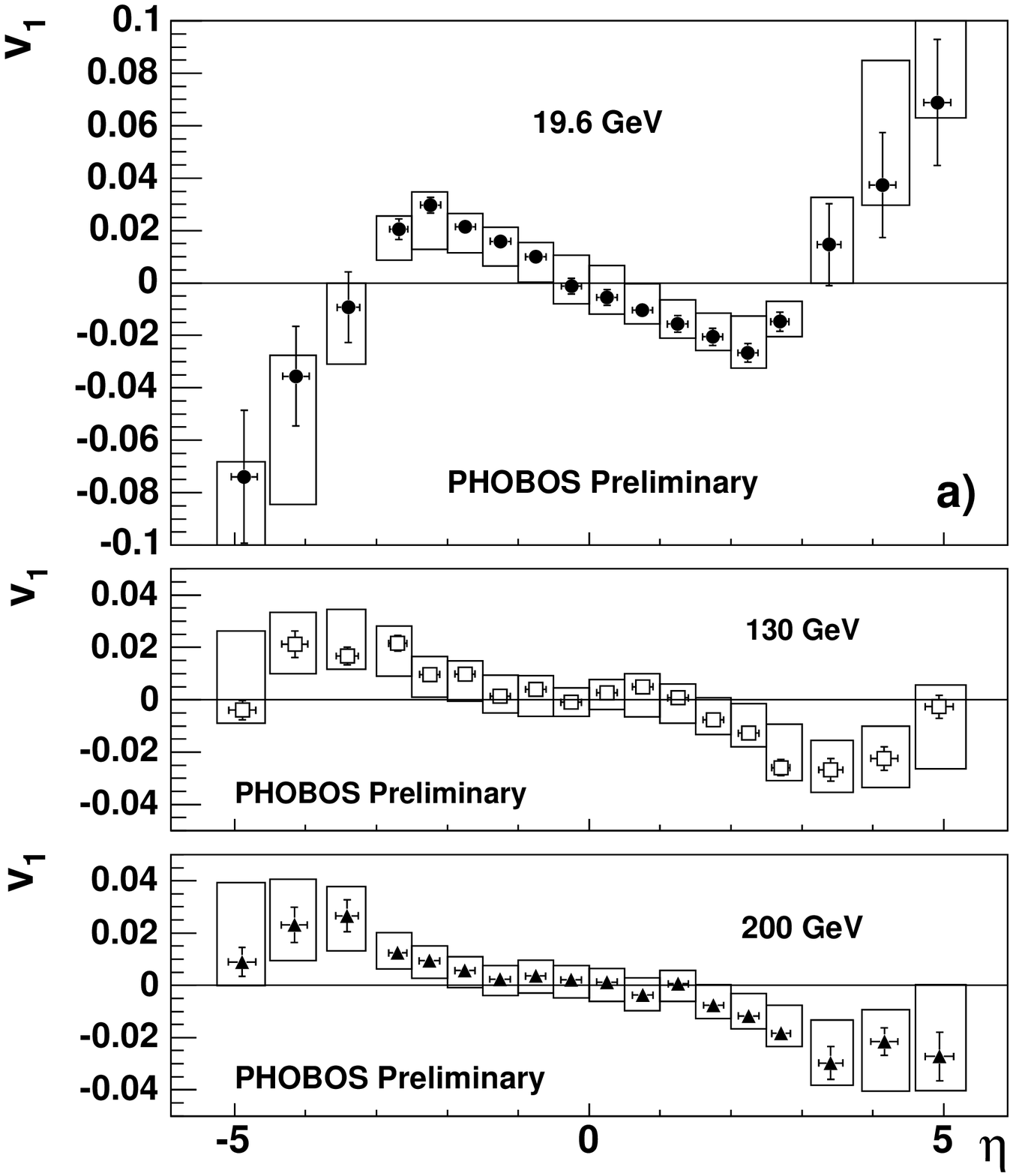}
\includegraphics[height=8.1cm]{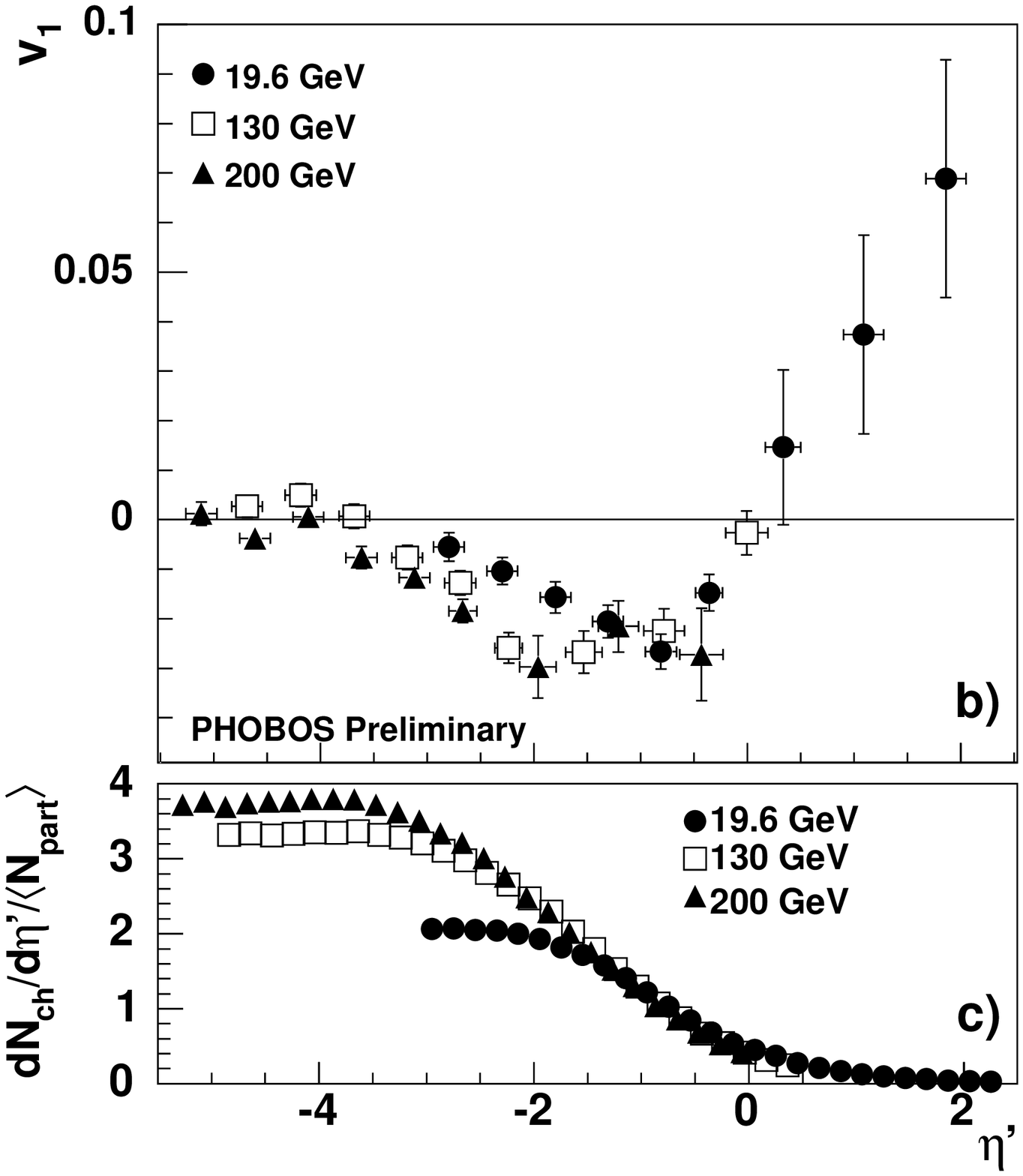}
\end{center}
\caption{\label{fig:v1}a) Directed flow measured as a function of $\eta$ for $\sqrt{s_{NN}}=$ 19.6 (top), 130(middle), and 200 GeV(bottom) Au+Au collisions. Error bars are statistical, the height of boxes represents the systematic error, and the width of the boxes marks the size of the $\eta$ bin. b) The same directed flow measurement as shown in a), but translated to $\eta '=\eta-y_{beam}$. Systematic error boxes are omitted for clarity. c) Scaled pseudorapidity distributions for 0-6\% central 19.6, 130, and 200 GeV Au+Au collisions as a function of $\eta '$\cite{limfrag}. Systematic errors not shown.}
\end{figure}

\section{Directed Flow}
Directed flow ($v_{1}$) is measured with a hit-based method. In this method, flow is calculated for hits measured in the octagon with an event plane 
determined from widely separated subevents (symmetric in $\eta$) in the rings. Subevents in the octagon are used for flow measured in the rings. Figure \ref{fig:v1}a) shows $v_{1}$ vs. $\eta$ for charged hadrons measured in Au+Au collisions at a range of energies. The $v_{1}$ values are averaged over a centrality range of 6-55\%. In the mid-rapidity region a significant change in slope can be seen from low to high energy Au+Au collisions. Similar directed flow results have been observed at low energy in NA49\cite{na49} and at high energy from STAR\cite{starv1}. In both 130 and 200 GeV measurements, we see that, within uncertainties, $v_{1}$ is flat at mid-$\eta$ and non-zero at high $|\eta|$, in contrast to the 19.6 GeV results.

A comparison of $v_{1}$ measured at different energies is best made by plotting $v_{1}$ against $\eta '=\eta-y_{beam}$. Figure \ref{fig:v1}b) shows the PHOBOS $v_{1}$ measurements for all three energies plotted as a function of $\eta '$.
The $v_{1}$ for the different energies is similar at $\eta '\stackrel{>}{\sim}-1.5$. It should be noted that the $v_{1}$ measurements at the low and high energies originate from different parts of the detector and have differing systematic error susceptibilities and sensitivities to the reaction plane. The agreement seen in the rest frame of the nucleus is reminiscent of the limiting fragmentation behavior observed in the particle density distributions shown in figure \ref{fig:v1}c). The same region in $\eta '$ shows agreement for all three energies in $v_{1}$ and $(dN_{ch}/d\eta')/<N_{part}>$. 

\section{Conclusions}
Charged hadron elliptic flow has been measured as a function of transverse momentum for 200 GeV Au+Au collisions. The centrality dependence of $v_{2}$ is shown over a large range in pseudorapidity. Peripheral $v_{2}$ as a function of $|\eta|$ is non-flat, even at midrapidity. The three measured centralities have the same overall $v_{2}(|\eta|)$ shape, differing only by a scale factor. PHOBOS measurements of charged hadron directed flow for 19.6, 130 and 200 GeV Au+Au collisions are shown over a large region of pseudorapidity. A change in slope of $v_{1}$ vs. $\eta$ at mid-rapidity can be seen from low to high energy Au+Au collisions. Looking in the rest frame of the nucleus, $v_{1}$ shows agreement at $\eta '\stackrel{>}{\sim}-1.5$ which is remininscent of limiting fragmentation in particle density distributions.

\ack{
\small
This work was partially supported by U.S. DOE grants 
DE-AC02-98CH10886,
DE-FG02-93ER40802, 
DE-FC02-94ER40818,  
DE-FG02-94ER40865, 
DE-FG02-99ER41099, and
W-31-109-ENG-38, by U.S. 
NSF grants 9603486, 
0072204,            
and 0245011,        
by Polish KBN grant 2-P03B-10323, and
by NSC of Taiwan under contract NSC 89-2112-M-008-024.}

\end{document}